\documentclass[12pt, a4paper]{article}

\usepackage[latin2]{inputenc}
\usepackage{graphicx}
\usepackage{amsmath}
\usepackage{fancyhdr}
\usepackage{authblk}

\title{Shannon entropy for imprecise and under-defined or over-defined information}
\date{}
\author{Vasile Patrascu \\ Tarom Information Technology,\\ 224F Bucurestilor Road, Otopeni, Rom\^ania,\\email: patrascu.v@gmail.com }

\begin{document}

\maketitle
\thispagestyle{fancy}
\begin{abstract}
 Shannon entropy was defined for probability 
distributions and then its using was expanded to measure the uncertainty 
of knowledge for systems with complete information. In this article, it 
is proposed to extend the using of Shannon entropy to under-defined or 
over-defined information systems. To be able to use Shannon entropy, the information is normalized by an affine transformation.
The construction of affine transformation is done in two stages: one for homothety  and another for translation. Moreover, the case of information with a certain degree of imprecision was included in this approach.
Besides, the article shows the using of Shannon entropy for some particular cases such as: 
neutrosophic information both in the trivalent and bivalent case, 
bifuzzy information, intuitionistic fuzzy information, imprecise fuzzy 
information, and fuzzy partitions.\\

{\bfseries\textit{Keywords}}: Shannon entropy, under-defined information, over-defined information, neutrosophic information, bifuzzy information, 
intuitionistic fuzzy information, imprecise fuzzy information, fuzzy partitions. 
\end{abstract}

\section{Introduction}

The Shannon entropy [\ref{r12}] plays an important role in the information uncertainty computing. Thus, if the information vector is defined by formula:

\begin{equation}\label{1.1}
p=(p_{1},p_{2},\ldots ,p_{n})\in\lbrack 0,1\rbrack ^{n} 
\end{equation}

and it verifies the condition of partition of unity, namely,

\begin{equation}\label{1.3}
\sum_{j=1}^{n}{p_{j}}=1 
\end{equation}

then, we compute the Shannon entropy using the well-known formula:

\begin{equation}\label{1.4}
E_{S}(p)=-\frac{1}{\ln (n)}\sum_{i=1}^{n}{p_{i}\ln (p_{i})} 
\end{equation}

The formula (\ref{1.4}) can be used only and only the information vector verifies the condition of the partition of unity (\ref{1.3}). But, what 
happens when the information is under-defined, when there exists the following inequality:

\begin{equation}\label{1.5}
\sum_{j=1}^{n}{p_{j}}<1 
\end{equation}

Also, what happens when the information is over-defined, when there exists the following inequality:

\begin{equation}\label{1.6}
\sum_{j=1}^{n}{p_{j}}>1 
\end{equation}

Usually, the degree of uncertainty for a vector information can have values in the interval $[0,1]$. Consequently, it is evidently that 
there exist different vectors from the $n$-dimensional unit hypercube that have the same value for the degree of uncertainty. For any value from 
the interval $[0,1]$ it can associated a class of vectors that have for the degree of uncertainty a specified value. Hence, it results the 
following idea: for each information vector $p$ that verifies the conditions (\ref{1.5}) or (\ref{1.6}), we must find an equivalent information vector $\hat{p}$ that verifies the condition (\ref{1.3}) and then we obtain the entropy $E_{S}(p)$ calculating entropy $E_{S}(\hat{p})$ using formula (\ref{1.4}). 
The obtaining of the equivalent vector $\hat{p}$ will be done using a normalization transformation and in the end it results a vector that verifies the condition of partition of unity (\ref{1.3}).

Usually, the equivalent vector $\hat{p}=(\hat{p}_{1},\hat{p}_{2},\ldots ,\hat{p}_{n})$ is obtained under the condition of the information proportionality, and 
it is determined a real and positive number $\lambda $, so that:

\begin{equation}\label{1.7}
\hat{p}=\lambda \cdot p 
\end{equation}

From the condition (\ref{1.3}) applied to vector $\hat{p}$ it results the number $\lambda $ : 

\begin{equation}\label{1.8}
\lambda =\left( \sum_{j=1}^{n}{p_{j}}\right) ^{-1} 
\end{equation}

Using the scaling factor, it is obtained the normalized vector $\hat{p}$:

\begin{equation}\label{1.9}
\hat{p}_{i}=\frac{p_{i}}{\sum_{j=1}^{n}{p_{j}}} 
\end{equation}

The formula  (\ref{1.9}) has a deficiency because it becomes instable when the sum of the components approaches to zero. In addition, we cannot use the 
normalization transformation defined by (\ref{1.9}), if the information vector $p=(p_{1},p_{2},\ldots ,p_{n})$ has a degree of imprecision defined by 
the parameter $s\in[0,1]$. In this context, we are faced to compute the Shannon entropy for the extended vector of information denoted by $P$ and defined by:

\begin{equation}\label{1.10}
P=(p_{1},p_{2},\ldots ,p_{n},s)
\end{equation}

It is observable that the formula (\ref{1.9}) does not take into account the degree of imprecision $s$ and this is an additional disadvantage. In order to 
solve the problem of information normalization, we will construct an affine transformation having two steps: a translation transformation [\ref{r8}] and a homothetic one [\ref{r7}]. Next, the article has the following structure: section 2 shows the construction of homothetic transformation; section 3 shows 
the construction of translation transformation; section 4 shows the aggregation of homothetic and translation in an affine transformation; section 5 
shows particular cases of using of the proposed entropy computing method; section 6 shows some conclusions while the last section is that of references.

\section{Homothetic transformation for over-defined information}

In this section, we will analyze the case of over-defined information and without having the degree of imprecision. In other words, the vector 
of information $p=(p_{1},p_{2},\ldots ,p_{n}) $ verifies the inequality (\ref{1.6}) and the degree of imprecision is zero, namely:

\begin{equation}\label{2.1}
\sum_{j=1}^{n}{p_{j}}>1 
\end{equation}

and

\begin{equation}\label{2.2}
 s=0 
\end{equation}

We can write the Jensen inequality [\ref{r5}], [\ref{r6}]:

\begin{equation}\label{2.4}
-\sum_{i=1}^{n}{p_{i} \ln (p_{i})}\le -\sum_{i=1}^{n}{p_{i}} \ln \left( \frac{1}{n}\sum_{j=1}^{n}{p_{j}}\right)  
\end{equation}

and the following equivalent forms:

\begin{equation}\label{2.5}
-\sum_{i=1}^{n}{\frac{p_{i}}{\sum_{j=1}^{n}{p_{j}}} \ln\left( \frac{p_{i}}{\sum_{j=1}^{n}{p_{j}}}\right) }\le \ln (n) 
\end{equation}

\begin{equation}\label{2.6}
-\frac{1}{\ln (n)}\sum_{i=1}^{n}{\frac{p_{i}}{\sum_{j=1}^{n}{p_{j}}}\ln 
\left( \frac{p_{i}}{\sum_{j=1}^{n}{p_{j}}}\right) }\le 1
\end{equation}

We obtained the Shannon entropy for over-defined information: 

\begin{equation}\label{2.7}
E_{S}(p)=-\frac{1}{\ln 
(n)}\sum_{i=1}^{n}{\frac{p_{i}}{\sum_{j=1}^{n}{p_{j}}} \ln 
\left( \frac{p_{i}}{\sum_{j=1}^{n}{p_{j}}}\right) }
\end{equation}

We will denote:

\begin{equation}\label{2.8}
\hat {p}_{i}=\frac{p_{i}}{\sum_{j=1}^{n}{p_{j}}}
\end{equation}

and (\ref{2.7}) becomes: 

\begin{equation}\label{2.9}
E_{S}(p)=-\frac{1}{ \ln (n)}\sum_{i=1}^{n}{\hat{p}_{i} \ln (\hat{p}_{i})}
\end{equation}

The formula (\ref{2.9}) represents the Shannon entropy that is utilized for the normalized information obtained using the homothetic transformation (\ref{2.8}). The 
information vector $\hat{p}$ describes normalized information and belongs to the polytope defined by (\ref{1.3}). But, the formula (\ref{2.8}) becomes quite 
instable when the sum $(\sum_{j=1}^{n}{p_{j}})$ is approaching zero. 
Because of that, we will directly use this normalization only when the sum $(\sum_{j=1}^{n}{p_{j}})$ is greater than one, namely when the 
information is over-defined. When the sum $(\sum_{j=1}^{n}{p_{j}})$ is less than one, namely the information is under-defined, we firstly do a 
translation and secondly the homothety defined by (\ref{2.8}). The translation is presented in the next section.

\section{Translation transformation for under-de\-fi\-ned information}

In the previous section, we have presented the normalization of over-defined information. This is reduced to a simple homothety [\ref{r7}]. 
As we have said earlier, the normalization of the under-defined information is done in two steps: a translation and then a homothety. We 
will construct the translation, starting from the assumption that two information vectors that have the same distances from the points with 
maximum certainty are equivalent and must have the same entropy or uncertainty. The points with maximum certainty are the vertices of the polytope described by (\ref{1.3}).
In other words, we will associate to each information vector $p$ describing an under-defined information, a vector that describes an 
over-defined information and keeps the distances from the vertices of the polytope defined by (\ref{1.3}). This condition ensures that we obtain an 
equivalent vector from the point of view of preserving the degree of uncertainty. In the next, we will consider two unit hypercubes: one in 
the $n$-dimensional space given by vector $p$ defined by (\ref{1.1}) and one in $(n+1)$-dimensional space given by vector $P$ defined by (\ref{1.10}).
The vertices of the polytope (\ref{1.3}) are the points where the Shannon entropy is zero and are described by vectors where a component is one 
and all the other ($n-1)$ components are zero. 

We consider an $n$-dimensional vector where the $j^{th}$ component  is $1$, namely:

\begin{equation}\label{3.1}
u=(0,\ldots ,0,1,0,\ldots ,0)
\end{equation}

and its extension in the $(n+1)$-dimensional space with zero on the last position for imprecision parameter $s$:

\begin{equation}\label{3.2}
U=(u,0)=(0,\ldots ,0,1,0,\ldots ,0,0)
\end{equation}

The vector obtained after the translation of vector $p$ will be defined by formula:

\begin{equation}\label{3.3}
\tilde{p}=(p_{1}+\vartheta ,p_{2}+\vartheta ,\ldots ,p_{n}+\vartheta )
\end{equation}

The translation parameter $\vartheta $ will be obtained solving the equation that preserves the distance:

\begin{equation}\label{3.4}
d(\tilde{p},u)=d(P,U)
\end{equation}

Using the Euclidean distance, the equation (\ref{3.4}) becomes:

\begin{equation}\label{3.5}
(p_{j}+\vartheta -1)^2+\sum_{\substack{i=1\\i\ne j}}^{n}{(p_{i}+\vartheta )^2}=(p_{j}-1)^2+\sum_{\substack{i=1\\i\ne j}}^{n}{p_{i}^2}+s^2 
\end{equation}

\begin{equation}\label{3.6}
n\vartheta ^{2}+2\vartheta (\sum_{j=1}^{n}{p_{j}}-1)-s^{2}=0
\end{equation}

We define the index of definedness by formula:

\begin{equation}\label{3.7}
\delta =\sum_{j=1}^{n}{p_{j}}-1
\end{equation}

and we obtain the following equation from (\ref{3.6}):

\begin{equation}\label{3.8}
 n\vartheta ^{2}+2\delta \vartheta -s^{2}=0
\end{equation}

Of course, there are two solutions:

\begin{equation}\label{3.9}
 \vartheta _{1,2}=\frac{-\delta \pm \sqrt{\delta ^{2}+ns^{2}}}{n}
\end{equation}

Since we are interested in over-defined information, we will only consider the variant with plus and it results for the translation parameter the following value:

\begin{equation}\label{3.10}
 \vartheta =\frac{-\delta +\sqrt{\delta ^{2}+ns^{2}}}{n}
 \end{equation}

It results the translated vector components:

\begin{equation}\label{3.11}
\tilde{p}_{i}=p_{i}+\frac{\sqrt{\delta ^{2}+ns^{2}}-\delta }{n}
\end{equation}

The translated vector $\tilde{p}$ represents over-defined information because it verifies the condition (\ref{1.6}), namely:

\begin{equation} \label{3.12}
\sum_{i=1}^{n}{\tilde{p}_{i}}=\sum_{i=1}^{n}{p_{i}}+n\vartheta
\end{equation}

\begin{equation}\label{3.13}
\sum_{i=1}^{n}{\tilde{p}_{i}}=\sum_{i=1}^{n}{p_{i}}-\delta +\sqrt{\delta^{2}+ns^{2}}
\end{equation}

\begin{equation}\label{3.14}
\sum_{i=1}^{n}{\tilde{p}_{i}}=1+\sqrt{\delta ^{2}+ns^{2}}\ge 1
\end{equation}

In the second step, because the vector information $\tilde{p}$ is over-defined, we can apply the homothetic transformation (\ref{2.8}) and at the end, it results the normalized vector $\hat{p}$.

\begin{equation}\label{3.15}
\hat{p}_{i}=\frac{\tilde{p}_{i}}{\sum_{j=1}^{n}{\tilde{p}_{j}}}
\end{equation}

\section{The affine transformation for information normalization}

After the presentation of the translation and homothetic transformations in the previous sections, we conclude that the normalized information vector is obtained applying an affine transformation $[\ref{r3}]$, $[\ref{r4}]$:

\begin{equation}\label{4.1}
 \hat{p}_{i}=\alpha p_{i}+\beta
\end{equation}

where the two parameters $(\alpha ,\beta )$ are defined by:

\begin{equation}\label{4.2}
\alpha =\frac{1}{1+\sqrt{\delta ^{2}+ns^{2}}}
\end{equation}

\begin{equation}\label{4.3}
\beta =\frac{\dfrac{\sqrt{\delta ^{2}+ns^{2}}-\delta}{n}}{1+\sqrt{\delta ^{2}+ns^{2}}}
\end{equation}

and after all it is obtained the following formula:

\begin{equation}\label{4.4}
\hat{p}_{i}=\frac{p_{i}+\dfrac{\sqrt{\delta ^{2}+ns^{2}}-\delta}{n}}{1+\sqrt{\delta ^{2}+ns^{2}}}
\end{equation}

In the next, we will consider the following two parameters:

the degree of under-definedness:

\begin{equation}\label{4.5}
u=max(-\delta ,0)
\end{equation}

the degree of over-definedness: 

\begin{equation}\label{4.6}
o=max(\delta ,0)
\end{equation}

Combining formulas (\ref{4.5}), (\ref{4.6}) and (\ref{4.4}) it results consequently:

\begin{equation}\label{4.7}
 \hat{p}_{i}=\frac{p_{i}+\dfrac{\sqrt{\delta^{2}+ns^{2}}-o+u}{n}}{1+\sqrt{\delta ^{2}+ns^{2}}} 
\end{equation}

\begin{equation}\label{4.8}
\hat {p}_{i}=\frac{p_{i}+\dfrac{2u}{n}+\dfrac{\sqrt{\delta^{2}+ns^{2}}-o-u}{n}}{1+\sqrt{\delta ^{2}+ns^{2}}}
\end{equation}

\begin{equation} \label{4.9}
\hat{p}_{i}=\frac{p_{i}+\dfrac{2u}{n}+\dfrac{\sqrt{\delta ^{2}+ns^{2}}-\vert\delta \vert }{n}}{1+\sqrt{\delta ^{2}+ns^{2}}}
\end{equation}

We define the cumulated imprecision:

\begin{equation}\label{4.10}
 h=\sqrt{\delta ^{2}+ns^{2}}-\vert \delta \vert
\end{equation}

As a final point, it results the formula for translated vector components:

\begin{equation}\label{4.11}
\hat {p}_{i}=\frac{p_{i}+\dfrac{2u+h}{n}}{1+\vert \delta \vert +h}
\end{equation}

After this, we compute the Shannon entropy for under-defined or over-defined information and supplementary having a degree of imprecision:

\begin{equation}\label{4.14}
 E_{S}(p)=-\frac{1}{\ln 
(n)}\sum_{i=1}^{n}{\left( \frac{p_{i}+\dfrac{2u+h}{n}}{1+\vert \delta \vert +h}\right) \ln \left( \frac{p_{i}+\dfrac{2u+h}{n}}{1+\vert \delta \vert +h}\right) }
\end{equation}

If the imprecision is zero, namely $s=0$, it results $h=0$ and one obtains the particular form:

\begin{equation}\label{4.15}
\hat {p}_{i}=\frac{p_{i}+\dfrac{2u}{n}}{1+\vert \delta \vert } 
\end{equation}

\begin{equation}\label{4.16}
 E_{S}(p)=-\frac{1}{\ln 
(n)}\sum_{i=1}^{n}{\left(\frac{p_{i}+\dfrac{2u}{n}}{1+\vert \delta \vert }\right)\ln \left( \frac{p_{i}+\dfrac{2u}{n}}{1+\vert \delta \vert }\right) } 
\end{equation}

In addition we can compute the Onicescu informational energy [\ref{r9}], the Tsallis entropy [\ref{r10}], [\ref{r16}] or Renyi entropy [\ref{r11}]:

Onicescu informational energy:

\begin{equation}\label{4.17}
E_{O}(p)=\sum_{i=1}^{n}{\left ( \frac{p_{i}+\dfrac{2u+h}{n}}{1+\vert \delta \vert +h}\right ) ^{2}}
\end{equation}

Tsallis entropy:

\begin{equation}\label{4.18}
E_{T}(p)=\frac{1-\sum_{i=1}^{n}{\left(\dfrac{p_{i}+\dfrac{2u+h}{n}}{1+\vert \delta \vert +h}\right)^{\alpha }}}{\alpha-1 }
\end{equation}

Renyi entropy:
\begin{equation}\label{4.19}
E_{R}(p)=\frac{1-\ln\left(\sum_{i=1}^{n}{\left(\dfrac{p_{i}+\dfrac{2u+h}{n}}{1+\vert \delta \vert +h}\right)^{\alpha }}\right)}{1-\alpha } 
\end{equation}
where $\alpha$ is a positive real number with $\alpha \ne 1$. When $\alpha\to 1$ the Tsallis and Renyi entropies recover the Shannon entropy.\\

\textbf{Observation.}
Usually, at practical level, we have $\delta \approx 0$ and we can take into account the following approximation for cumulated imprecision:

\begin{equation} \label{4.12}
\sqrt{\delta ^{2}+ns^{2}}-\vert \delta \vert \approx s\sqrt{n}
\end{equation}

It is obtained:

\begin{equation}\label{4.13}
\hat {p}_{i}\approx \frac{p_{i}+\dfrac{2u+s\sqrt{n}}{n}}{1+\vert \delta \vert +s\sqrt{n}}
\end{equation}
On the other hand, (\ref{4.13}) is the exact formula for imprecise and complete information.

\section{Some particular cases for Shannon entropy}
In the following we will present some particular cases for using of the Shannon entropy: neutrosophic information, 
bifuzzy information, intuitionistic fuzzy information, imprecise fuzzy information, and fuzzy partitions.

\subsection{Three-valued Shannon entropy for neutrosophic information.}

The neutrosophic information proposed by Smarandache [\ref{r14}], [\ref{r15}] is defined by three parameters: degree of truth $T\in [0,1]$, degree 
of falsity $F\in [0,1]$ and degree of neutrality $I\in [0,1]$. The vector $p=(T,I,F)$ represents the primary information.
We define the neutrosophic definedness and under-definedness by following two formulas:

\begin{equation}\label{5.1}
D=T+F+I-1 
\end{equation}

\begin{equation}\label{5.2}
U=\max (-D,0)
\end{equation}

If $D<0$ then the neutrosophic information is under-defined and if $D>0$ then the neutrosophic information is over-defined.
In this case for three-valued Shannon entropy, we consider three points where the certainty is maximum, namely $p_{T}=(1,0,0)$, $p_{I}=(0,1,0)$ and $ p_{F}=(0,0,1)$.
 Using (\ref{4.15}) it results the three-valued normalized information $\hat{p}=(\hat{T},\hat{I},\hat{F})$:

\begin{equation}\label{5.3}
\hat{T}=\frac{T+\dfrac{2U}{3}}{1+\vert D\vert } 
\end{equation}

\begin{equation}\label{5.4}
\hat{I}=\frac{I+\dfrac{2U}{3}}{1+\vert D\vert }
\end{equation}

\begin{equation}\label{5.5}
\hat{F}=\frac{F+\dfrac{2U}{3}}{1+\vert D\vert } 
\end{equation}

The neutrosophic information $(\hat{T},\hat{I},\hat{F})$ verifies the condition of the partition of unity:

\begin{equation}\label{5.6}
\hat{T}+\hat{I}+\hat{F}=1
\end{equation}

The Shannon entropy is calculated using formula (\ref{4.16}) and it results:

\begin{equation}\label{5.7}
\begin{split}
E_{S}(p)=&-\frac{\left(\dfrac{T+\dfrac{2U}{3}}{1+\vert D\vert }\right)  \ln\left( \dfrac{T+\dfrac{2U}{3}}{1+\vert D\vert }\right)}{\ln(3)}-   \frac{\left(\dfrac{I+\dfrac{2U}{3}}{1+\vert D\vert }\right) \ln\left( \dfrac{I+\dfrac{2U}{3}}{1+\vert D\vert }\right)}{\ln (3)}- \\
         &\quad \frac{\left(\dfrac{F+\dfrac{2U}{3}}{1+\vert D\vert }\right) \ln\left( \dfrac{F+\dfrac{2U}{3}}{1+\vert D\vert }\right)}{\ln (3)}
\end{split}
\end{equation}

\subsection{Bi-valued Shannon entropy for neutrosophic information}

The neutrosophic information is described by parameters: degree of truth $ \mu \in [0,1]$, degree of falsity $\nu \in [0,1]$ and degree of imprecision $\omega \in [0,1]$.

Whe define the following parameters:

the bifuzzy definedness:

\begin{equation}\label{6.1}
\delta =\mu +\nu -1
 \end{equation}

the bifuzzy incompleteness:

\begin{equation}\label{6.2}
\pi =\max (-\delta ,0)
\end{equation}

the cumulated imprecision:

\begin{equation}\label{6.3}
h=\sqrt{\delta ^{2}+2\omega ^{2}}-\vert \delta \vert
\end{equation}

In this case for bi-valued Shannon entropy, we consider two points where the certainty is maximum, namely $p_{T}=(1,0,0)$ and $ p_{F}=(0,1,0)$.
 It results its equivalent fuzzy degree of truth $\hat{\mu }$ and its fuzzy degree of falsity $\hat{\nu }$: 

\begin{equation}\label{6.4}
\hat{\mu }=\frac{\mu +\pi +\dfrac{h}{2}}{1+\vert \delta \vert +h} 
\end{equation}

\begin{equation}\label{6.5}
\hat{\nu }=\frac{\nu +\pi +\dfrac{h}{2}}{1+\vert \delta \vert +h}
\end{equation}

The fuzzy information $\hat{p}=(\hat{\mu },\hat{\nu })$ represents the bi-valued normalized form of the primary information $p=(\mu ,\nu ,\omega )$ and there exists the equality:

\begin{equation}\label{6.6}
\hat{\mu }+\hat{\nu }=1
\end{equation}

Using the fuzzy information $\hat{p}=(\hat{\mu },\hat{\nu })$ that was associated to the neutrosophic information $p=(\mu ,\nu ,\omega )$ we will 
compute the bi-valued Shannon entropy by the following formula:

\begin{equation}\label{6.7}
\begin{split}
E_{S}(p)=&-\frac{\left( \dfrac{\mu +\pi +\dfrac{h}{2}}{1+\vert \delta \vert +h}\right)  \ln\left( \dfrac{\mu +\pi +\dfrac{h}{2}}{1+\vert \delta \vert +h}\right)}{\ln (2)} -\\
    	&\quad \frac{\left( \dfrac{ \nu +\pi +\dfrac{h}{2}}{1+\vert \delta \vert +h}\right)  \ln\left( \dfrac{\nu +\pi +\dfrac{h}{2}}{1+\vert \delta \vert +h}\right) }{\ln (2)}
\end{split}
\end{equation}

\subsection{Shannon entropy for bifuzzy information}

The bifuzzy information [\ref{r1}], [\ref{r2}] is described by two parameters: 
degree of truth $\mu \in [0,1]$ and degree of falsity $\nu \in [0,1]$.
We define the bifuzzy definedness $\delta \in [-1,1]$ and bifuzzy incompleteness  $\pi\in [0,1]$ by:

\begin{equation}\label{7.1}
\delta =\mu +\nu -1
 \end{equation}

\begin{equation}\label{7.2}
\pi =\max (-\delta ,0)
\end{equation}

In this case we consider two points where the certainty is maximum, namely $p_{T}=(1,0)$ and $p_{F}=(0,1)$. We compute the fuzzy 
degree of truth $\hat{\mu}$ and fuzzy degree of falsity $\hat{\nu}$ using (\ref{4.15}) and it results:

\begin{equation}\label{7.3}
\hat{\mu }=\frac{\mu +\pi }{1+\vert \delta \vert }
\end{equation}

\begin{equation}\label{7.4}
\hat{\nu }=\frac{\nu +\pi }{1+\vert \delta \vert }
\end{equation}

There exists the equality:

\begin{equation}\label{7.5}
\hat{\mu }+\hat{\nu }=1 
\end{equation}

Using the associated fuzzy information $\hat{p}=(\hat{\mu},\hat{\nu})$ to the bifuzzy information $p=(\mu ,\nu )$ we will compute the Shannon entropy by the following formula derived from (\ref{4.16}):

\begin{equation}\label{7.6}
E_{S}(p)=-\frac{\left( \dfrac{\mu +\pi }{1+\vert \delta \vert }\right) \ln \left( \dfrac{\mu +\pi }{1+\vert \delta \vert }\right) +\left( \dfrac{\nu +\pi }{1+\vert 
\delta \vert }\right) \ln \left( \dfrac{\nu +\pi }{1+\vert \delta \vert }\right) }{\ln (2)} 
\end{equation}

\subsection{Shannon entropy for intuitionistic fuzzy information}

The intuitionistic fuzzy information [\ref{r1}], [\ref{r2}] is described by two parameters: degree of truth $\mu \in[0,1]$ and degree of falsity $
\nu \in[0,1]$ verifying the following inequality $1\ge \mu +\nu $.

We define the degree of incompleteness $\pi$ by:

\begin{equation}\label{8.1}
 \pi =1-\mu -\nu
 \end{equation}

The information is under-defined or incomplete and we will associate the following fuzzy information with degree of truth $\hat{\mu }$ and degree of falsity $\hat{\nu }$:

\begin{equation}\label{8.2}
\hat {\mu }=\frac{\mu +\pi }{1+\pi } 
\end{equation}

\begin{equation}\label{8.3}
\hat {\nu }=\frac{\nu +\pi }{1+\pi }
\end{equation}

with:

\begin{equation}\label{8.4}
\hat {\mu }+\hat{\nu }=1 
\end{equation}
Using the associated fuzzy information $\hat{p}=(\hat{\mu },\hat{\nu })$ to the intuitionistic fuzzy information $p=(\mu ,\nu )$, we will compute the Shannon entropy by the following formula derived from (\ref{4.16}):

\begin{equation}\label{8.5}
E_{S}(p)=-\frac{\left( \dfrac{\mu +\pi }{1+\pi }\right) \ln \left( \dfrac{\mu +\pi }{1+\pi 
}\right) +\left( \dfrac{\nu +\pi }{1+\pi }\right) \ln \left( \dfrac{\nu +\pi }{1+\pi }\right) }{\ln (2)} 
\end{equation}

Equivalent with:

\begin{equation}\label{8.6}
E_{S}(p)=-\frac{\left(\dfrac{\bar{\mu }}{\bar{\mu }+\bar{\nu }}\right) \ln\left( \dfrac{\bar{\mu }}{\bar{\mu }+\bar{\nu }}\right) +\left( \dfrac{\bar{\nu }}{\bar{\mu 
}+\bar{\nu }}\right) \ln \left( \dfrac{\bar{\nu }}{\bar{\mu }+\bar{\nu }}\right) }{\ln (2)}
\end{equation}

where the negation is calculated using formula:

\begin{equation}\label{8.7}
 \bar{x }=1-x
\end{equation}

\subsection{Shannon entropy for imprecise fuzzy information}

The fuzzy information [\ref{r17}] is described by the degree of truth $\mu\in [0,1]$ while the imprecise fuzzy information is described by the pair $
p=(\mu ,\sigma )$, where $\mu \in[0,1]$ is the degree of truth and $\sigma \in\left[ 0,\dfrac{1}{2}\right] $ is the degree of imprecision. We 
must mention that $\nu =1-\mu $ represents the degree of falsity. The imprecise fuzzy information can be seen as particular neutrosophic case 
where $(T,I,F)$ are defined by:
$$T=\mu $$ 
$$I=2\sigma$$ 
$$F=1-\mu$$

In this framework, it results the following particular values for definedness $\delta $, cumulated imprecision $h$, fuzzy degree of truth $\hat{\mu }$
and fuzzy degree of falsity $\hat{\nu}$:

\begin{equation}\label{9.1}
 \delta =\mu +\nu -1=0
\end{equation}

\begin{equation}\label{9.2}
 h=2\sigma \sqrt{2}
\end{equation}

\begin{equation}\label{9.3}
 \hat{\mu }=\frac{\mu +\sigma \sqrt{2}}{1+2\sigma \sqrt{2}}
\end{equation}

\begin{equation}\label{9.4}
\hat {\nu }=\frac{\nu +\sigma \sqrt{2}}{1+2\sigma \sqrt{2}}
\end{equation}

with:

\begin{equation}\label{9.5}
\hat {\mu }+\hat{\nu }=1 
\end{equation}

Using (\ref{4.16}), we obtain Shannon entropy for imprecise fuzzy information:

\begin{equation}\label{9.6}
E_{S}(p)=-\frac{\left( \dfrac{\mu +\sigma \sqrt{2}}{1+2\sigma \sqrt{2}}\right) \ln \left( \dfrac{\mu +\sigma \sqrt{2}}{1+2\sigma \sqrt{2}}\right) +\left( \dfrac{\nu +\sigma \sqrt{2}}{1+2\sigma \sqrt{2}}\right) \ln \left( \dfrac{\nu +\sigma \sqrt{2}}{1+2\sigma \sqrt{2}}\right) }{\ln (2)}
\end{equation}

\subsection{ Bi-valued Shannon entropy for fuzzy partition}

We consider the fuzzy partition $(w_{1},w_{2},\ldots ,w_{n})$ and there exists the equality,

\begin{equation}\label{10.1}
 w_{1}+w_{2}+\ldots +w_{n}=1
\end{equation}

We order the membership functions and get the following decreasing set of values:

\begin{equation}\label{10.2}
 o_{1}\ge o_{2}\ge \ldots \ge o_{n} 
\end{equation}

where

\begin{equation}\label{10.3}
 o_{1}=max(w_{1},w_{2},\ldots ,w_{n})
\end{equation}

and

\begin{equation}\label{10.4}
 o_{n}=min(w_{1},w_{2},\ldots ,w_{n}) 
\end{equation}

Firstly, we construct an intuitionistic fuzzy representation where $\mu =o_{1}$, $\nu =o_{2}$ and $\pi =1-o_{1}-o_{2}$. Secondly, we 
construct the fuzzy representation where $\hat{\mu }$ and $\hat{\nu } $ are defined by (\ref{8.2}) and (\ref{8.3}). It results:

\begin{equation}\label{10.5}
 \hat{ \mu}=\frac{o_{1}+\pi }{1+\pi }
\end{equation}

\begin{equation}\label{10.6}
 \hat{\nu}=\frac{o_{2}+\pi }{1+\pi } 
\end{equation}

Using formula (\ref{4.16}) for associated fuzzy information $(\hat{\mu },\hat{\nu })$, one obtains the bi-valued Shannon entropy for the fuzzy partition $w$:

\begin{equation}\label{10.7}
E_{S}(w)=-\frac{\left( \dfrac{o_{1}+\pi }{1+\pi }\right) \ln \left( \dfrac{o_{1}+\pi }{1+\pi }\right) +\left( \dfrac{o_{2}+\pi }{1+\pi }\right) \ln \left( \dfrac{o_{2}+\pi }{1+\pi }\right) }{\ln (2)}
\end{equation}

with its equivalent form derived from (\ref{8.6}):

\begin{equation}\label{10.8}
E_{S}(w)=-\frac{\left( \dfrac{\bar{o_{1}}}{\bar{o_{1}}+\bar{o}_{2}}\right) \ln 
\left( \dfrac{\bar{o}_{1}}{\bar{o_{1}}+\bar{o}_{2}}\right) +\left( \dfrac{\bar{o}_{2}}{\bar{o_{1}}+\bar{o}_{2}}\right) \ln 
\left( \dfrac{\bar{o}_{2}}{\bar{o_{1}}+\bar{o}_{2}}\right) }{\ln (2)}
\end{equation}

There are other non-logarithmic formulas for bi-valued fuzzy partition entropy computing such as the following three:

\begin{equation}\label{10.9}
 E_{K}(w)=1-\frac{\vert o_{1}-o_{2}\vert }{1+\pi } 
\end{equation}

\begin{equation} \label{10.10}
E_{E}(w)=\sqrt{\frac{1-2o_{1}+\sum_{i=1}^{n}{o_{i}^{2}}}{1-2o_{2}+\sum_{i=1}^{n}{o_{i}^{2}}}}
 \end{equation}

\begin{equation}\label{10.11}
 E_{P}(w)=\frac{1-o_{1}}{1-o_{2}}=\frac{\bar{o_{1}}}{\bar{o}_{2}} 
\end{equation}

\section{Conclusion}

The article presents a method of using Shannon entropy for under-defined or over-defined information with a certain degree of imprecision. For 
this purpose, a two-step normalization procedure is proposed: a translation and a homothetic one. After the presentation, the procedure 
is used for calculating Shannon's entropy in the case of particular representations of information such as neutrosophic information, bifuzzy 
information, intuitionistic fuzzy information, imprecise fuzzy information and fuzzy partitions. In the case of neutrosophic 
information, two variants are possible: the first is the trivalent variant in which the certainty has three prototypes: true, neutral and 
false; the second is the bivalent variant in which the certainty has two prototypes: true and false. The article mentions that the presented 
method of normalization can be used for other formulas such as Onicescu information energy, Tsallis entropy or Renyi entropy.

\section*{References}

\begin{enumerate}
\item\label{r1} K. T. Atanassov, Intuitionistic fuzzy sets. Fuzzy Sets Syst. 20, 87-96, 1986.
\item\label{r2} K. T. Atanassov, Intuitionistic Fuzzy Sets: Theory and Applications. Studies in Fuzziness and Soft Computing, vol. 35, Physica-Verlag, Heidelberg ,1999.
\item\label{r3} M. Berger, Geometry I, Berlin, Springer, ISBN 3-540-11658-3, 1987.
\item\label{r4} M. Hazewinkel, Affine transformation, Enciclopedia of Mathematics, Springer, ISBN 978-1-75608-010-4, 2001.
\item\label{r5} M. Hazewinkel, Encyclopedia of Mathematics, Springer, ISBN 978-1-55608-010-4, 2001.
\item\label{r6} J. L. W. V. Jensen, Sur les fonctions convexes et les inegalites entre les valeurs moyennes, Acta Mathematica, 30 (1): 175-193, doi:10.1007 / BF02418571, 1906.
\item\label{r7} B. Meserve, Homothetic transformations, Fundamental Concept of Geometry, Addison-Wesley, pp. 166-169, 1955.
\item\label{r8} W. Osgood, W. Graustein, Plane and solid analytic geometry, The Macmillan Company, P330, 1921.
\item\label{r9} O. Onicescu, Energie informationnelle, Comptes Rendus Hebdomadaires des Sciances de l'Academie des Sciences,  Serie A 263 , 841-842, 1966.
\item\label{r10} C. Tsallis, Possible generalization of Boltzmann-Gibbs statistics, J. Stat. Phys., 52, 479-487, 1988.
\item\label{r11} A. Renyi, On Measures of Entropy and Information. Proceedings of the Fourth Berkeley Symposium on Mathematical Statistics and Probability, 
Volume 1: Contributions to the Theory of Statistics, 547-561, University of California Press, Berkeley, Calif., 1961.
\item\label{r12} C. E. Shannon, A mathematical theory of communication, Bell System Tech. J. 27, 379-423, 1948.
\item\label{r14} F. Smarandache, A Unifying Field in Logics: Neutrosophic Logic, Multiple valued logic, An international journal, 8, no. 3, 385-438, 2002.
\item\label{r15} F. Smarandache, Neutrosophic Set - A Generalization of the Intuitionistic Fuzzy Set, International Journal of Pure and Applied Mathematics,24, no. 3, 287-297, 2005.
\item\label{r16} J. Harvda, F. Charvat, Quantification method of classification processes-concept of structural $\beta$-entropy, Kybernetika (Prague) 3, 30-35, 1967.
\item\label{r17} L. A. Zadeh, Fuzzy sets, Information Control 8, 338-353, 1965.

\end{enumerate}

\end{document}